\documentclass[11pt]{article}
\begin{document}
\input amssym.def
\date{}
\title{Towards Adelic Noncommutative Quantum Mechanics}

\author{G. S. Djordjevi\'c and Lj. Ne\v si\'c\\
Department of Physics, Faculty of Science\\
 P.O. Box 224, 18000 Ni\v s, Serbia and Montenegro\\
 e-mail: gorandj@junis.ni.ac.yu}

\maketitle

\begin{abstract}
A motivation of using noncommutative and nonarchimedean
geometry on very short distances is given.
 Besides some mathematical
preliminaries, we give a short introduction in adelic quantum
mechanics. We also recall
to basic ideas and tools embedded in $q$-deformed and
noncommutative quantum mechanics. A rather fundamental
approach, called deformation quantization, is noted.
A few relations between noncommutativity and nonarchimedean
spaces as well as similarities between corresponding quantum
theories on them are pointed out. An extended Moyal product
in a proposed form of adelic noncommutative
quantum mechanics is considered.
We suggest some question for
future investigations.
\end{abstract}

\section{Introduction}

It is widely accepted that standard picture of space-time should
be changed around and beyound Planck scale. ``Measuring'' of spacetime
geometry under distances smaller than Planck length $l_p$ is not
accesible even to Gedanken experiments. It serves an idea of
``quantization'' and "discretization" of spacetime and a natural cutoff
when using a quantum field theory to describe related phenomena.
We are pointing out two possibilties for a reasonable mathematical background
of a quantum theory on very small distances, The {\it first} one comes from
the idea of of spacetime coordinates as noncommuting operators
\begin{equation}
\label{djo-noncom}
[\hat x^{i},\hat x^{j}]=i\theta^{ij}.
\end{equation}
Some noncommutativity of configuration space should not be a surprise in
physics since quantum phase space with the canonical commutation relation
\begin{equation}
\label{djo-com}
[\hat x^i,\hat k^j]=i\hbar\delta^{ij} ,
\end{equation}
where $x^i$ are coordinates and $k^j$ are the
corresponding momenta,
is the well-known example of noncommutative (pointless) geometry.
This relation is connected in a natural way with the uncertainty principle
and a ``fuzzy'' spacetime pictures at distances $~\theta^{1/2}$.
Altough, it seems to have good physical sense for $\theta^{1/2}\sim l_p$,
characteristic noncommutative distances could be related to gauge couplings
\cite{djo-wess}, closer to observable distances. It should be noted that
deriving of uncertainty relation ($\delta x>l_p$) leads to an ``strange''
notion of quantum line and probably beyound archimedean geometry, because
a coordinate always commute with itself \cite{djo-av}!

The {\it second} promising approach to the physics at the
Planck scale, based on non-archimedean geometry was suggested \cite{djo-volo}.
The simplest way to describe such a geometry (oftenly called also
ultrametric) is by using $p$-adic number fields
$\Bbb Q_p$ ($p$ is a prime).
On the basis of the Ostrovski theorem \cite{djo-vladi} there is
no other nontrivial possibilities (besides
field of real numbers $\Bbb R$) to complete field of rational numbers $\Bbb Q$
in respect to a (nontrivial) norm. Remind
that all physical numerical experimental data belong $\Bbb Q$.

There have been many interesting applications of $p$-adic
numbers and non-Archimedean geometry in various parts of modern theoretical
and mathematical physics
(for a review, see \cite{djo-vladi,djo-breke}).
However we restrict ourselves here mainly to $p$-adic \cite{djo-vladi2}
and adelic \cite{djo-mentor} quantum mechanics (QM). It should be noted that
 adelic QM have appeared quite useful in
quantum cosmology. The appearance of space-time discreteness in adelic
formalism (see, e.g. \cite{djo-mentormi}), as well as in
noncommutative QM, is an encouragement for the further
investigations. We emphasize the role of Feynman's
$p$-adic path integral method on nonarchimedean spaces.

The $p$-adic analysis and noncommutativity also play a role in some areas of
"macroscopic" physics \cite{djo-jackiw}. We list shortly a few of
similarities between non-Archimedean and  noncommutative structures them and
discuss in more details a new observed relation between an ordering on
commutative ring in frame of deformation quantization \cite{djo-waldmann} and
an ordering on $p$-adic spaces in an intention to develop path
integration on $p$-adics by "slicening" of trajectories \cite{djo-zele}.

%We start this paper with a short introduction to $p$-adic numbers, adeles and their functions. After that, $p$-adic
%and adelic QM based on the Weyl quantization and Feynman's
%path integral are presented. We continue with a recall on q-deformation,
%noncommutative QM and deformation quantization and emphasize
%some relation of them with the nonarchimedean approach.
%Finally, we discuss a possibilty to formulate noncommutative
%adelic QM and some aspects of the extended Moyal product
%\cite{djo-kijev}.

\section{$p$-Adic numbers and adeles}

Any $x\in \Bbb Q_p$ can be presented in the form \cite{djo-vladi}
\begin{equation}
\label{djo-canonexp}
x = p^\nu(x_0+x_1p+x_2p^2+\cdots)\ ,\quad \nu\in \Bbb Z,
\end{equation}
where $x_i = 0,1,\cdots,p-1$ are digits. $p$-Adic norm of any term
$x_ip^{\nu+i}$ in the canonical expansion (\ref{djo-canonexp}) is
$|x_ip^{\nu+i}|_p
=p^{-(\nu+i)}$ and the strong triangle inequality
holds,
 {\it i.e.} $|a+b|_p\leq\hbox{max}\{|a|_p,|b|_p\}$. It follows that $|x|_p = p^{-\nu}$ if
$x_0\neq 0$.
Derivatives of $p$-adic valued functions $\varphi : \Bbb Q_p\to \Bbb Q_p$ are
defined as in the real case, but with respect to the
$p$-adic norm. There is no integral $\int\varphi(x)dx$ in a sense of
the Lebesgue measure \cite{djo-vladi}, but one can introduce
$\int^b_a\varphi(x)dx
= \Phi(b)-\Phi(a)$ as a functional of analytic functions
$\varphi(x)$, where $\Phi(x)$ is an antiderivative of $\varphi(x)$.
In
the case of map $f:\Bbb Q_p\to \Bbb C$ there is
well-defined Haar measure. One can use the Gauss integral
\begin{equation}
\int_{\Bbb Q_\upsilon}\chi_\upsilon(ax^2+bx)dx =
\lambda_\upsilon(a)|2a|^{-{1\over2}}_\upsilon{\chi}_v
\big(-{b^2\over4a}\big)\ ,\quad a\not=0,\ \upsilon =\infty,2,3,5,\cdots\!,
\end{equation}
where index $\upsilon$ denotes real ($\upsilon =\infty$) and $p$-adic cases,
$\chi_\upsilon$ is an additive character: $\chi_\infty(x)=\exp(-2\pi ix)$,
$\chi_p(x) = \exp(2\pi i\{x\}_p)$,\quad  where $\{x\}_p$
is the fractional part of $x\in \Bbb Q_p$. $\lambda_\upsilon (a)$ is
the complex-valued arithmetic function \cite{djo-vladi}.
An adele \cite{djo-geljfand} is an infinite sequence $a=(a_\infty,
a_2,..., a_p,...)$, where
$a_\infty\in \Bbb R\equiv \Bbb Q_\infty$, $a_{p}\in \Bbb Q_{p}$ with a restriction
that $a_{p} \in \Bbb Z_{p}$ for all but a finite set $S$ of primes
$p$. The set of all adeles $\Bbb A$  may be  regarded as a subset of
direct topological product $\Bbb Q_\infty\times\prod_p \Bbb Q_p$.
$\Bbb A$ is a topological space, and can be considered as a ring with
respect to the componentwise addition and multiplication. An elementary function on
adelic ring $\Bbb A$ is
\begin{equation}
\label{djo-elementary}
        \varphi (x)=\varphi _{\infty}(x_{\infty})\prod_{p}^{}\varphi
_{p}(x_{p})
        =\prod_{v}^{} \varphi _{v}(x_{v})  \;\;
\end{equation}
with the main restriction that
$\varphi (x)$ must satisfy $\varphi_{p}(x_{p})=\Omega
(|x_{p}|_{p})$
for all but a finite number of $p$.
Characteristic function on
$p$-adic integers $\Bbb Z_p$
is defined by $\Omega (|x|_p) =1, \ 0\leq |x|_p \leq 1$ and
$\Omega (|x|_p) =0, \ |x|_p>1$.

The Fourier transform of the characteristic
function (vacuum state) $\Omega(|x_p|)$ is $\Omega(|k_p|)$.
All finite linear combinations of elementary functions
(\ref{djo-elementary}) make  the set $\cal D(\Bbb A)$ of the Schwartz-Bruhat
functions. The Hilbert space $L_2(\Bbb A)$ is a space of complex-valued functions
$\psi_1(x)$, $\psi_2(x),\dots$, with the scalar product and norm.

\section{Adelic quantum mechanics}

In foundations of standard QM one
usually starts with a representation of the canonical commutation
relation (\ref{djo-com}).
In formulation of $p$-adic QM
\cite{djo-vladi2} the multiplication
$\hat q\psi\rightarrow x \psi$ has no meaning for $x\in\Bbb Q_p$
and $\psi(x)\in \Bbb C$. In the real case momentum and hamiltonian
are infinitesimal generators of space and time translations, but, since
$\Bbb Q_{p}$ is disconnected field, these
infinitesimal transformations become meaningless.
However, finite transformations remain meaningful and the corresponding
Weyl and evolution operators are $p$-adically well defined.

Canonical commutation relation (\ref{djo-com}) in $p$-adic case can
be represented by the Weyl operators ($h=1$)
\begin{equation}
\hat Q_p(\alpha) \psi_p(x)=\chi_p(\alpha x)\psi_p(x)
\end{equation}
\begin{equation}
\hat K_p(\beta)\psi(x)=\psi_p(x+\beta) .
\end{equation}
\begin{equation}
\hat Q_p(\alpha)\hat K_p(\beta)=\chi_p(\alpha\beta)\hat K_p(\beta)\hat Q_p(\alpha).
\end{equation}
It is possible to introduce the family of unitary operators
\begin{equation}
\hat W_p(z)=\chi_p(-\frac 1 2 qk)\hat K_p(\beta)\hat Q_p(\alpha), \quad
z\in\Bbb Q_p\times\Bbb Q_p,
\end{equation}
that is a unitary representation of the Heisenberg-Weyl group.
Recall that this group consists of the elements $(z,\alpha)$ with the
group product
$(z,\alpha)\cdot(z',\alpha ')=(z+z',\alpha+\alpha '+\frac 1 2 B(z,z'))$,
where $B(z,z')=-kq'+qk'$ is a skew-symmetric bilinear
form on the phase space.
Dynamics of a $p$-adic quantum model is described by a unitary
operator of evolution $U(t)$ formulated in terms of its kernel
$K_t(x,y)$,
$U_p(t)\psi(x)=\int_{\Bbb Q_p}K_t(x,y)\psi(y) dy$. In this way
\cite{djo-vladi2} $p$-adic QM is given by
a triple $(L_2(\Bbb Q_p), W_p(z_p), U_p(t_p))$.

Keeping in mind that standard QM can be also given as the
corresponding triple, ordinary and $p$-adic
QM can be unified in
the form of adelic QM \cite{djo-mentor}
\begin{equation}
(L_2(A), W(z), U(t)).
\end{equation}
$L_{2}(\Bbb A)$ is the Hilbert space on $\Bbb A$, $W(z)$ is a unitary
representation of the Heisenberg-Weyl group on $L_2(\Bbb A)$ and
$U(t)$ is a  unitary representation of the
evolution operator on $L_2(\Bbb A)$.
The evolution operator $U(t)$ is defined by
\begin{equation}
U(t)\psi(x)=\int_{\Bbb A} K_t(x,y)\psi(y)dy=\prod\limits_{v}{}
\int_{\Bbb Q_{v}}K_{t}^{(v)}(x_{v},y_{v})\psi^{(v)}(y_v) dy_{v}.
\end{equation}
Note that any adelic eigenfunction  has the form
\begin{equation}
\label{djo-psi}
\Psi(x) =
\Psi_\infty(x_\infty)\prod_{p\in S}\Psi_p(x_p)
\prod_{p\not\in S}\Omega(|x_p|_p) , \quad x\in \Bbb A,
\end{equation}
where $\Psi_{\infty}\in L_2(\Bbb R)$,
$\Psi_{p}\in L_2(\Bbb Q_p)$.
In the low-energy limit adelic
QM becomes ordinary one.

A suitable way to calculate propagator in $p$-adic QM is by
$p$-adic generalization of Feynman's path integral \cite{djo-vladi2}. There is
no natural ordering on $\Bbb Q_p$. However,
a bijective continuous map $\varphi$ from from the set of
$p$-adic numbers $\Bbb Q_p$ to the subset $\varphi(\Bbb Q_p)$ of real numbers
$\Bbb R$ \cite{djo-vladi} was proposed. This map can be defined by (for an older injective
version see \cite{djo-zele})
\begin{equation}
\label{map}
\varphi(x)=|x|_p \sum_{k=0}^\infty x_k p^{-2k}.
\end{equation}
Than, a linear order on $\Bbb Q_p$ is given by the following
definition: $x<y$ if $|x|_p<|y|_p$ or when
$|x|_p =|y|_p$ there exists such index $m\geq0$ that
digits satisfy $x_0 = y_0, x_1 = y_1, \cdots,x_{m-1} = y_{m-1}\ ,x_m<y_m$.
One can say: $\varphi(x) > \varphi(y)$ iff $x>y$.

In the case of
harmonic oscillator \cite{djo-zele}, it was shown that there exists the limit
\begin{eqnarray}
K_p(x^{\prime\prime},t^{\prime\prime};x^\prime,t^\prime)=
\lim_{n\to\infty}
K_p^{(n)}(x^{\prime\prime},t^{\prime\prime};x^\prime,t^\prime) =
\lim_{n\to\infty}N^{(n)}_p(t^{\prime\prime},t^\prime) \nonumber \\
\times\int_{\Bbb Q_{p}}\cdots
\int_{{\Bbb Q}_{p}}
\chi_p\bigg(-{1\over h}\sum^n_{i=1}\bar S(
q_i,t_i;q_{i-1},t_{i-1})\bigg)dq_1\cdots dq_{n-1}\ ,
\end{eqnarray}
where $N^{(n)}_p(t^{\prime\prime},t^\prime)$ is the corresponding
normalization factor for the harmonic oscillator.
The subdivision of $p$-adic time segment $t_0<t_1<\cdots<t_{n-1}<t_n$
is made
according to linear order on $\Bbb Q_p$.
In a similar way we have calculated
path integrals for a few quantum models. For the references see \cite{djo-kijev}.
Moreover, we were able to obtain general expression for the propagator of the
systems with quadratic action  (for the details
see  \cite{djo-general}), without ordering
\begin{equation}
\label{djo-quadratic}
K_p(x^{\prime\prime}\!,t^{\prime\prime}\!;x^\prime\!,\!t^\prime)=
\!\lambda_p
\!\bigg(\!-\frac{1}{2h}{\partial^2\bar S\over \partial
x^{\prime\prime}\partial x^\prime}
\bigg)\!
\Big\arrowvert\frac{1}{h}{\partial^2\bar S\over \partial
x^{\prime\prime}\partial x^\prime}
\Big\arrowvert^{\frac{1}{2}}_p
\chi_p
\!\bigg(\!-{1\over h}\bar
S(x^{\prime\prime}\!,t^{\prime\prime}\!;x^\prime\!,\!t^\prime)
\bigg)\!.
\end{equation}
Replacing an index $p$
with $v$ in (\ref{djo-quadratic}) we can write quantum-mechanical amplitude $K$ in ordinary and
all $p$-adic
cases in the same, compact (and adelic) form.

\section{Relations between noncommutative and $p$-adic QM}

Noncommutative geometry is geometry which is described by an associative algebra
${\cal A}$ whici is usually noncommutative and in which the set of points,
if it exists at all, is relegated to a secondary role.
Noncommutative spaces have arised in investigation of brane configurations in string and
M-theory. Since the one-particle sector of field theories leads to QM, a study of this topic has attracted much of interest. For single particle
QM, the corresponding Heisenberg algebra is needed. In addition
to (\ref{djo-noncom}) and (\ref{djo-com}) one choose
\begin{equation}
\label{djo-mom}
[p^i,p^j]=i\Phi^{ij}
\end{equation}
There are a lot of possibilities in choosing $\theta$ and $\Phi$. Although one can
take $\theta^{ij}$ and $\Phi^{ij}$ are antisymetric nonconstant tensors (matrices),
often, the simplest nontrivial case is considered: $\theta^{ij}=const$ and
$\Phi^{ij}=0$. Another realization of noncommutativity is possible by $q$-deformation
of a space, for example, {\it Manin plane} $xy=qyx$ and $q$-deformed "classical" phase
space $px=qpx$. This approach leads to
to latticelike (discrete) structure of space-time \cite{djo-wess1}.

A field $\Psi(x)$ as a function of the noncommuting coordinates $x$ can be
used as Schro\"dinger wave function obeying the free field equation.
Other realization,
based on star product ($v\ast \Psi$) instead of standard multiplication
($V\cdot \Psi$) of a potential and wave funaction have been considered in corresponding Schr\"odinger equation
too (i.e. see \cite{djo-mezincesku}).

The passage from one level of physical theory to more refined another,
using what mathematicians call deformation theory is nothing extraordinary new.
In a similar way, there is an old idea that QM is some kind of
deformed classical mechanics. For a review see \cite{djo-sternheimer}.
In fact, deformation quantization is closely related to Weyl quantization,
shortly sketched in the previous section.

One direction of the investigation led to Moyal bracket and
Moyal (star) product, widely used now in noncommutative QM
\begin{equation}
%\label
f\ast_m g=\chi_\infty
\left(
-\frac{h}{8\pi^2}P
\right)
(f,g)=
fg+\sum_{r=1}^{\infty}
\left(
\frac{ih}{4\pi}
\right)^r P^r(f,g).
\end{equation}
Several integral formulas have been introduced for the star
product and an (formal) parameter of deformation is finally
related to some form of Planck constant $h$. Quantisation
can be taken as a deformation of the standard
associative and commutative product, now called a star product,
of classical observables driven by the
Poissone bracket $P$.
By the intuition, classical mechanics is understood as the limit
QM when $h\rightarrow 0$.

Some connections between  $p$-adic analysis and quantum
deformations has been noticed during
the last ten years.
It has been observed that the Haar measure on
$SU_q(2)$ coincides with the Haar measure on the field of $p$-adic
numbers $\Bbb Q_p$ if $q=p^{-1}$ \cite{djo-av}.
There is a potential such that the spectrum of the $p$-adic Schr\"odinger-like
(diffusion) equation
\cite{djo-vladi}
\begin{equation}
\label{djo-povisilica}
D\psi (x)+V(|x|_p)\psi (x)=E\psi (x)
\end{equation}
is the same one as in the case of $q$-deformed oscillator found by
Biedenharn \cite{djo-b} for$q=1/p$.
For more details see \cite{djo-av}.
%Recently \cite{djo-trnovo}, it has been proposed a new pseudodifferential
%operator with rational part of $p$-adic numbers $\{x\}_p$. It will be an
%interesting question to investigate a possible relation between spectra
%of this operator and $q$-deformed models.

In a develop of the representation theory for the star
product algebras in deformation quantization some non-Archimedean behavior
is noted\cite{djo-waldmann}. We find this relation very intrigue
and discuss it in more details. Recall,
that a star product $\ast $, on a Poisson manifold (M,$\pi$) is a (formal)
associative $\Bbb C[[\lambda]]$ - bilinear product for
$C^{\infty}(M)[[\lambda]]$
\begin{equation}
f\ast g = \sum_{r=0}^{\infty} \lambda^r C_r(f,g),
\end{equation}
with bidiffererential operators $C_r$.
$\Bbb C[[\lambda]]$ is an algebra of
{\it formal} (in a formal parameter $\lambda$) series
$\sum_{r=0}^{\infty} \lambda^r C_r, \ C_r\in \Bbb C$
(a convergence of this series is still not under a
consideration). $C^{\infty}(M)[[\lambda]]$ is the space of formal
series of smooth functions ($x\in M,\ f_r(x)\in C^{\infty}(M)$), for a
fixed but variable x. With interpretation $\lambda \leftrightarrow
\hbar$ we identify $C^{\infty}(M)[[\lambda]]$ with the algebra of
observables of the quantum system corresponding to $(M,\pi)$.
Let $R$ be on an ordered (commutative associative unital) ring $R$
\cite{djo-waldmann},
Let us note, that a concept of ordered ring is neccesary one wish
to define the positive
functionals on a $C^\ast$ algebra ($C=R(i)=\{a+bi,\ a,b\in R\}$). It is related
with Gelfand-Naimark's theorem on commutative spaces. By means of the
positive functionals on $C^\ast$ algebras we can
reconstruct the (points on) ``starting'' manifold, that one on which
an algebra of complex function will give the ``original''  $C^\ast$
algebra. This is a motivation for generalization on noncommutative
space. In this case, the corresponding product of formal
functions will be the $\ast$ product. Now, $R[[\lambda]]$
will be an algebra of formal series on the ordered ring $R$. Then,
if $R$ is an ordered ring, $R[[\lambda]]$ will be ordered in a
canonical way, too, by the definition
\begin{equation}
\sum_{r=r_0}^{\infty} \lambda^ra_r>0,\ \ if \ \ a_{r_0}>0.
\end{equation}
In other words, a formal series in $R[[\lambda]]$ will be a
``positive`` one, if first nonzero coeficient (an element of ordered
ring $R$ is positive).
\noindent It should be noted that the concept of ordered ring fits
naturally to formal power series and thus to Gerstenhaber's deformation
theory \cite{djo-gerstenhaber}. Then $R[[\lambda]]$ will be {\it non-Archimedean}.
For example, if we take that $a_0=-1$, and $a_1=n$ ($n\in \Bbb N$,
because for any commutative ordered ring with unit, set of integers
is embedded in $R$, $\Bbb N\subset R$, all others coeficients
can be zero, we have $-1+n\lambda<0$ or $n\lambda<1$ for all
$n\in \Bbb Z$.
The interpretation in formal theory is that the deformation parameter
$\lambda$ is ``very small'' compared to the other numbers in R.
We see that Waldmann's definition of the ordered algebra of formal
series on ordered ring $R$ immediately leads to the non-archimedean
``structure''.
It could be a good indication to use ultrametric spaces and p-adics
when physical deformation parameter is very small.
We would like to underline that Zelanov's ordering by means of map
(\ref{map}) $\Bbb Q_p$, i.e. on normed, ultrametric algebra, in frame of
$p$-adic QM for $\lambda=1/p^2$ is related
to the ordering of formal series on ordered rings
in the frame of deformation quantization!

%Finally, we mention: Kozyrev proved \cite{djo-kozyrev}
%the space of generalized functions over the noncommutative plane
%is isomorphic to a Gelfand triple to the space of generalized
%functions over a $p$-adic disc, or, one can say the
%noncommutative plane is equivalent to a $p$-adic disc; Connes showed that a
%spectral interpretation of zeros of zeta-function described by a quantum chaotic
%model could be done by the construction of noncommutative spaces of adelic
%classes \cite{djo-connes}.

\section{The adelic Moyal product and noncommutative QM}

The presented connections noncommutative vs.
"nonarchimedean" QM
suggest a need to formulate a quantum theory that may connect as
much as possible nonarchimedean and noncommutative effects and
structures. At the present level of quantum theory
on adeles, a formulation of Noncommutative Adelic QM
seems as the most prommising attempt.

An enough simple frame for that might be
representation of an algebra of operators (\ref{djo-noncom}),
(\ref{djo-com}) and (\ref{djo-mom}). It could be done by
linear transformations on
corresponding simplectic structure and deformed and extended
bilinear product $B$. Correpondance between classical functions
and quantum operators would be provided by Weyl quantisation. An equivalent
formulation of noncommutative adelic QM by the
triple $(L_2(A_\theta), W_\theta(z), U_\theta(t))$, does not
seem to have principles obstacles. In this approach an adele of
coordinates $x_A$ would be replaced by
a serie of noncommutative operators
$\hat x_A$,
where adelic properties of corresponding
eigenvalues is still "conserved".

Now, we has to consider
a $p$-adic and adelic generalization of the Moyal product.
Let us consider classical space with coordinates
$x^1,x^2$, $\cdots,x^D$. Let $f(x)$ be a classical function
$f(x)=f(x^1,x^2,\cdots,x^D)$.
Then, with the respect to the Fourier transformations and the usual Weyl
quantization, we have
\begin{equation}
\hat f(x)=\int_{\Bbb Q_\infty^D} dk \
\chi_\infty(-k \hat x) \tilde f(k)\equiv  f(\hat x).
\end{equation}
Let us now have two classical functions $f(x)$ and $g(x)$ and we are
interested in operator product $\hat f(x) \hat g(x)$. In the real case this
operator product is
\begin{equation}
\label{djo-rmoyal}
(\hat f \cdot \hat g)(x)=\int \int dk dk' \ \chi_\infty(-k\hat x)
\chi_\infty(-k' \hat x) \tilde f(k)\tilde g(k').
\end{equation}
Using the Baker-Campbell-Hausdorff formula, the relation (\ref{djo-noncom})
and then the coordinate representation one finds
the  Moyal product in the form
\begin{equation}
(f\ast g)(x)=\int_{\Bbb Q_p^D}\int_{\Bbb Q_p^D} dk dk' \ \chi_\upsilon\left ( -(k+k')x+\frac 1 2
k_ik'_j\theta^{ij}\right )\tilde f(k)\tilde g(k').
\end{equation}
Note that
we already used our generalization from  $\Bbb Q_\infty$ to $\Bbb Q_\upsilon$.
In the real case we  use
$k_i\rightarrow -(i/2\pi)(\partial/\partial x^i)$ and obtain the well
known form
$(f\ast g)(x)=\chi_\infty\left(-\frac{\theta^{ij}}{2(2\pi)^2}
\frac{\partial}{\partial y^i}\frac{\partial}{\partial z^j}\right ) f(y)g(z)|_{y=z=x}$.
In the $p$-adic case such an straightforward generalization is not possible (but, some
kind of psudodifferentiaton could be useful).
Thus, as the $p$-adic Moyal product we take
\begin{equation}
(f \ast g)(x)=\int_{\Bbb Q_p^D}\int_{\Bbb Q_p^D} dk dk'
\ \chi_p(-(x^ik_i+x^jk'_j)+\frac{1}{2} k_ik'_j\theta^{ij})\tilde f(k)\tilde g(k').
\end{equation}
We can writedown the adelic Moyal product of "classical" adelic functions
$f_A=(f_\infty,f_2,...,f_p,...)$, $g_A=(g_\infty,g_2,...,g_p,...)$ on
$\Bbb R\times \prod_{p\in S} \Bbb Q_p$ $\times \prod_{p\not\in S}\Bbb Z_p$ space
\begin{eqnarray}
(f \ast g)(x)=\int_{\Bbb A^D}\int_{\Bbb A^D} dk dk'
\ \chi(-(x^ik_i+x^jk'_j)+\frac{1}{2} k_ik'_j\theta^{ij})\tilde f_A(k)\tilde g_A(k')
\end{eqnarray}
Taking into account (\ref{djo-elementary}), (\ref{djo-rmoyal}) and the property
of the Fourier transform of $\Omega$ function, one has
\begin{eqnarray}
(f \ast g)(x)=\chi_\infty\left(-\frac{\theta^{ij}}{2(2\pi)^2}
\frac{\partial}{\partial y^i}\frac{\partial}{\partial z^j}\right ) f(y)g(z)|_{y=z=x}\nonumber \\
\times\prod_{p\in S} \int_{\Bbb Q_p^D}\int_{\Bbb Q_p^D} dk dk'
\ \chi_p(-(x^ik_i+x^jk'_j)+\frac{1}{2} k_ik'_j\theta^{ij})\tilde f_p(k)\tilde g_p(k')\nonumber \\
\times \prod_{p\notin S}\int_{\Bbb Z_p^D}\int_{\Bbb Z_p^D} dk dk'
\ \chi_p(-(x^ik_i+x^jk'_j)+\frac{1}{2} k_ik'_j\theta^{ij}).
\end{eqnarray}
It can be shown
that if for all $p$,
$\varphi(x)=\Omega(x)$, the adelic Moyal product becomes real one.

\section*{Acknowledgments}
%{\bf{Acknowledgments}}
G.Dj. is partially supported by DFG Project
``Noncommutative space-time structure - Cooperation with Balkan Countries''.
We would like to thank P. Aschieri, I. Bakovic, B. Jurco and
S. Waldmann for useful discussions.

\end{document}